
\magnification 1200
\newbox\leftpage \newdimen\fullhsize \newdimen\hstitle \newdimen\hsbody
\tolerance=1000\hfuzz=2pt
\def\bigans{b }
\def\answ{b }
%
\ifx\answ\bigans\message{(This will come out unreduced.}
\magnification=1200\baselineskip=16pt plus 2pt minus 1pt
\hsbody=\hsize \hstitle=\hsize 
\else\def\apans{l }\message{ lyman or hepl (l/h) (lowercase]) ? }
\read-1 to \apansw\message{(This will be reduced.}
\let\lr=L
\magnification=1000\baselineskip=16pt plus 2pt minus 1pt
\voffset=-.31truein\vsize=7truein
\hstitle=8truein\hsbody=4.75truein\fullhsize=10truein\hsize=\hsbody
\ifx\apansw\apans\special{ps: landscape}\hoffset=-.59truein
  \else\hoffset=.05truein\fi
\output={\ifnum\pageno=0 
  \shipout\vbox{\hbox to \fullhsize{\hfill\pagebody\hfill}}\advancepageno
  \else
  \almostshipout{\leftline{\vbox{\pagebody\makefootline}}}\advancepageno
  \fi}
\def\almostshipout#1{\if L\lr \count1=1
      \global\setbox\leftpage=#1 \global\let\lr=R
  \else \count1=2
    \shipout\vbox{\ifx\apansw\apans
    \special{ps: landscape}\fi 
      \hbox to\fullhsize{\box\leftpage\hfil#1}}  \global\let\lr=L\fi}
\fi
%
\catcode`\@=11 
\newcount\yearltd\yearltd=\year\advance\yearltd by -1900

%
%
\def\draftmode{\def\draftdate{{\rm preliminary draft:
\number\month/\number\day/\number\yearltd\ \ \hourmin}}%
\headline={\hfil\draftdate}
\writelabels\baselineskip=10pt plus 2pt minus 2pt
{\count255=\time\divide\count255 by 60 \xdef\hourmin{\number\count255}
        \multiply\count255 by-60\advance\count255 by\time
   \xdef\hourmin{\hourmin:\ifnum\count255<10 0\fi\the\count255}}}

\def\nolabels{\def\eqnlabel##1{}\def\eqlabel##1{}\def\reflabel##1{}}

 \def\writelabels{\def\eqnlabel##1{%
 {\escapechar=` \hfill\rlap{\hskip.09in\string##1}}}%
 \def\eqlabel##1{{\escapechar=` \rlap{\hskip.09in\string##1}}}%
 \def\reflabel##1{\noexpand\llap{\string\string\string##1\hskip.31in}}}

\nolabels
%
\global\newcount\secno \global\secno=0
\global\newcount\meqno \global\meqno=1
\def\newsec#1{\global\advance\secno by1\message{(\the\secno. #1)}
\xdef\secsym{\the\secno.}\global\meqno=1
\bigbreak\bigskip
\noindent{\bf\the\secno. #1}\par\nobreak\medskip\nobreak}
\xdef\secsym{}
\def\appendix#1#2{\global\meqno=1\xdef\secsym{\hbox{#1.}}\bigbreak\bigskip
\noindent{\bf Appendix #1. #2}\par\nobreak\medskip\nobreak}
%
%
\def\eqnn#1{\xdef #1{(\secsym\the\meqno)}%
\global\advance\meqno by1\eqnlabel#1}
\def\eqna#1{\xdef #1##1{\hbox{$(\secsym\the\meqno##1)$}}%
\global\advance\meqno by1\eqnlabel{#1$\{\}$}}
\def\eqn#1#2{\xdef #1{(\secsym\the\meqno)}\global\advance\meqno by1%
$$#2\eqno#1\eqlabel#1$$}
%
\newskip\footskip\footskip14pt plus 1pt
minus 1pt 
\def\f@@t{\baselineskip\footskip\bgroup\aftergroup\@foot\let\next}
\setbox\strutbox=\hbox{\vrule height9.5pt depth4.5pt width0pt}
\global\newcount\ftno \global\ftno=0
\def\foot{\global\advance\ftno by1\footnote{$^{\the\ftno}$}}
%
%
\global\newcount\refno \global\refno=1
\newwrite\rfile
\def\ref{\nref}
\def\nref#1{\xdef#1{[\the\refno]}\ifnum\refno=1\immediate
\openout\rfile=refs.tmp\fi\global\advance\refno by1\chardef\wfile=\rfile
\immediate\write\rfile{\noexpand\item{#1\ }\reflabel{#1}\pctsign}\findarg}
\def\findarg#1#{\begingroup\obeylines\newlinechar=`\^^M\pass@rg}
{\obeylines\gdef\pass@rg#1{\writ@line\relax #1^^M\hbox{}^^M}%
\gdef\writ@line#1^^M{\expandafter\toks0\expandafter{\striprel@x #1}%
\edef\next{\the\toks0}
\ifx\next\em@rk\let\next=\endgroup\else\ifx\next\empty%
\else
\immediate\write\wfile{\the\toks0}\fi\let\next=\writ@line\fi\next\relax}}
\def\striprel@x#1{} \def\em@rk{\hbox{}} {\catcode`\%=12\xdef\pctsign{
\def\semi{;\hfil\break}
\def\addref#1{\immediate\write\rfile{\noexpand\item{}#1}} 
\def\listrefs{\immediate\closeout\rfile
\baselineskip=14pt\centerline{{\bf References}}\bigskip{\frenchspacing%
\escapechar=` \input refs.tmp\vfill\eject}\nonfrenchspacing}
\def\startrefs#1{\immediate\openout\rfile=refs.tmp\refno=#1}
\def\figures{\centerline{{\bf Figure Captions}}\medskip\parindent=40pt}
\def\fig#1#2{\medskip\item{Fig.~#1:  }#2}
\catcode`\@=12 
%
%
\def\noblackbox{\overfullrule=0pt}
\hyphenation{anom-aly anom-alies coun-ter-term coun-ter-terms}
\def\inv{^{\raise.15ex\hbox{${\scriptscriptstyle -}$}\kern-.05em 1}}
\def\dup{^{\vphantom{1}}}
\def\Dsl{\,\raise.15ex\hbox{/}
\mkern-13.5mu D} 
\def\dsl{\raise.15ex\hbox{/}\kern-.57em\partial}
\def\ksl{\raise.15ex\hbox{/}\kern-.57em k}
\def\del{\partial}
\def\Psl{\dsl}
\def\tr{{\rm tr}} \def\Tr{{\rm Tr}}
\def\lspace{\ifx\answ\bigans{}\else\qquad\fi}
\def\lbspace{\ifx\answ\bigans{}\else\hskip-.2in\fi} 
\def\boxeqn#1{\vcenter{\vbox{\hrule\hbox{\vrule\kern3pt\vbox{\kern3pt
        \hbox{${\displaystyle #1}$}\kern3pt}\kern3pt\vrule}\hrule}}}
\def\mbox#1#2{\vcenter{\hrule \hbox{\vrule height#2in
                \kern#1in \vrule} \hrule}}  
%
\def\CAG{{\cal A/\cal G}}   
\def\CA{{\cal A}} \def\CC{{\cal C}} \def\CF{{\cal F}} \def\CG{{\cal G}}
\def\CL{{\cal L}} \def\CH{{\cal H}} \def\CI{{\cal I}} \def\CU{{\cal U}}
\def\CB{{\cal B}} \def\CR{{\cal R}} \def\CD{{\cal D}} \def\CT{{\cal T}}
\def\e#1{{\rm e}^{^{\textstyle#1}}}
\def\grad#1{\,\nabla\]_{{#1}}\,}
\def\gradgrad#1#2{\,\nabla\]_{{#1}}\nabla\]_{{#2}}\,}
\def\ph{\varphi}
\def\psibar{\overline\psi}
\def\om#1#2{\omega^{#1}{}_{#2}}
\def\vev#1{\langle #1 \rangle}
\def\lform{\hbox{$\sqcup$}\llap{\hbox{$\sqcap$}}}
\def\darr#1{\raise1.5ex\hbox{$\leftrightarrow$}\mkern-16.5mu #1}
\def\lie{\hbox{\it\$}} 
\def\ha{{1\over2}}
\def\half{{\textstyle{1\over2}}} 
\def\roughly#1{\raise.3ex\hbox{$#1$\kern-.75em\lower1ex\hbox{$\sim$}}}
\ref\adler{S. L. Adler, {\sl Ann. Phys.} {\bf 67} (1971) 599.}
\ref\drummond{I. J. Drummond and S.J. Hathrell, {\sl Phys. Rev.} {\bf D22
} (1980) 343.}
\ref\hawking{
S. W. Hawking and G. F. R. Ellis, {\sl The Large Scale
Structure of the Space--Time}, Cambridge Univ. Press, Cambridge 1973.}
\ref\tarrach{R. Tarrach, {\sl Phys. Lett.} {\bf B133} (1983) 259.}
\ref\barton{G. Barton, {\sl Phys. Lett.} {\bf B237} (1990) 559.}
\ref\scharnhorst{K. Scharnhorst,
 {\sl Phys. Lett.} {\bf B236} (1990) 354.}
\ref\berestetskii,{V. B. Berestetskii,
 E. M. Lifshitz and L. P.
Pitaevskii, {\sl Quantum Electrodynamics}, $2^{nd}$ edition, Pergamon
Press, Oxford 1982.}

\ref\daniels{R. D. Daniels and G. M. Shore, {\sl SWAT-93-9, hep-th
9310114.}}
\ref\ben{S. Ben--Menahem, {\sl Phys. Lett.} {\bf B250} (1990) 133.}
\ref\bartonscharnhorst{ G. Barton and K. Scharnhorst,
  {\sl J. Phys.} {\bf A26} (1993) 2037.}
\ref\birrel{R. D. Birrel and P. C. W. Davies, {\sl Quantum Fields in
Curved Space}, Cambridge Univ. Press, Cambridge 1982.}
\ref\pascual{
P. Pascual, J. Taron and R. Tarrach, {\sl Phys. Rev.} {\bf D39}
(1989) 2993.}
\ref\pisarski{R. Pisarski, {\sl Phys. Rev. Lett.}
{\bf 63} (1989) 1129
\semi
R. Kobes, G. Kunstatter and A. Rebhan, {\sl Phys. Rev. Lett.} {\bf
64} (1990) 2992.}
\ref\weldon{H. A. Weldon, {\sl Phys. Rev.} {\bf D28} (1983) 2007.}
\ref\gribosky{P. S. Gribosky and B. R. Holstein, {\sl Zeitsch. Phys.}
 {\bf C47} (1990) 205.}

\ref\bedaque{P. F. Bedaque and A. Das, {\sl Phys. Rev.} {\bf D45} (1992)
2906.}
\ref\braaten{E. Braaten and R. Pisarski, {\sl Nucl. Phys.} {\bf
B337} (1990) 569.}
\ref\niemi{A. J. Niemi and G. W. Semenoff, {\sl Ann. Phys. }
 {\bf 152} (
1984) 105; {\sl Nucl. Phys.} {\bf B230} (1984) 181\semi
I. Ojima, {\sl Ann. Phys.} {\bf 137} (1981) 1\semi
H. Umezawa, H. Matsumoto and M. Tachiki, {\sl Thermofield Dynamics and
Condensed States}, North Holland, Amsterdam 1982.}
\ref\dolan{L. Dolan and R. Jackiw, {\sl Phys. Rev.} {\bf D9} (1974) 3357.}
\ref\bordag{M. Bordag, D. Robaschik and E. Wieczorek, {\sl Ann. Phys.}
{\bf 165} (1985) 192.}

\ref\candelas{ P. Candelas, {\sl Ann. Phys.} {\bf 143} (1982) 241.}

\def\ii{\'{\char'20}}
\null
\vfill
\centerline{\bf SPEED OF LIGHT IN NON--TRIVIAL VACUA}
\bigskip \bigskip
\centerline{\bf
Jos\'e I. Latorre\footnote{*}{{\rm bitnet : }{\tt latorre@ebubecm1}},
Pedro Pascual and
Rolf Tarrach\footnote{**}{{\rm bitnet : }{\tt
rolf@ebubecm1}}} \bigskip
\centerline{\it Departament d'Estructura i Constituents de la Mat\`eria}
\centerline{\it Facultat de F\ii sica, Universitat de Barcelona}
\centerline{\it Diagonal 647, 08028-Barcelona\ \ SPAIN}
\centerline{ and \ \ {\sl IFAE}}

\vfill
\centerline{\bf Abstract}

We unify all existing results on the change of the speed of
low--energy photons  due
to  modifications of the vacuum, finding that  it is
 given by a universal constant times the quotient
of the difference of energy densities between the usual and modified
vacua over the mass of the electron to the fourth power.
 Whether photons move faster or slower than $c$
depends only
on the lower or higher energy density of the modified
vacuum, respectively. Physically, a higher energy density is
 characterized by the presence of additional
   particles (real or virtual) in the vacuum
whereas a lower one stems from the absence of some virtual modes.
 We then
carry out a
systematic study of the speed of propagation of massless
particles
 for several field theories up to two loops on a thermal
vacuum.  Only low--energy massless particles corresponding to a massive
theory show genuine modifications of their speed while remaining
massless.  All other modifications are mass-related, or running
mass-related.  We also develop a formalism for the Casimir vacuum which
parallels the thermal one and check that photons travel faster than $c$
between plates.

 {\baselineskip=13pt \vfill \noindent{\bf UB-ECM-PF 94
\hfill } \eject

\newsec{Review of existing results}
\bigskip
The usual vacuum in a Quantum Field Theory is a state characterized by
the absence of real particles and classical fields and by its
Minkowskian geometry. Electromagnetic and gravitational fields as well
as massless particles propagate though it with the same constant,
Lorentz invariant speed, $c$. When the vacuum is modified,
so is the speed of propagation of particles and fields.
This quantum field theoretical effect has been analyzed within QED for
low--energy photons in several cases that we now review.

  Adler first studied the propagation of photons in an anisotropic
vacuum given by an external constant uniform magnetic field, $\vec B$,
and
obtained the result ($\hbar=c=\epsilon_0=1$) \adler
\eqn\adlerone{\eqalign{&
v_{\parallel}=
 1- {8 \over 45} \alpha^2 {\vec B^2\over
m_e^4} \sin^2\theta\qquad < 1\,,\cr
&v_{\perp}=
 1- {14\over 45} \alpha^2 {\vec B^2\over
m_e^4} \sin^2\theta\qquad < 1\,,\cr}}
for polarizations respectively coplanar with and perpendicular to the
plane defined by $\vec B$ and the direction of propagation. Here,
$m_e$ is the mass of the electron, $\alpha$ is the fine structure
constant and $\theta$ is the angle between $\vec B$ and the direction of
propagation.

For a (homogeneous and isotropic) Robertson-Walker gravitational
background with Friedmann cosmology, Drummond and Hathrell obtained
\drummond
\eqn\edrummond{v=1+{11\over 45} \alpha G_N {\rho+p\over m_e^2}\qquad
>1\,,}
where $G_N$ is Newton's constant and $\rho$ and $p$ are the energy
density and pressure (which satisfy by  the energy condition
\hawking,
$\rho+p>0$). Notice that the speed of light is now greater than $c$.

For a homogeneous and isotropic thermal vacuum, one of us first computed
the change in the speed of light \tarrach.
 That result was corrected in Ref. \barton\ and
is reobtained here in Sect. 5 with the result
\eqn\erolf{
v=1- {44 \pi^2\over 2025}\alpha^2 {T^4\over m_e^4}\qquad <1\,
,}
where $T$ is the temperature and $k_B=1$.

Finally, Scharnhorst \scharnhorst\
 and, independently, Barton \barton\
 worked out the
propagation of
light in a (anisotropic and non-translational invariant) Casimir vacuum
with the result
\eqn\escharnhorst
{v=1+{11\pi^2\over 8100}\alpha^2 {1\over a^4 m_e^4}
\cos^2\theta\qquad>1\,,}
where $a$ is the distance between plates and $\theta$ is the angle
between the direction of propagation and the normal to the plates.
Again, photons propagate faster than $c$ in this case.

There are several features worth mentioning in these four expressions.
First, we note that the electron mass plays the role of an infrared
regulator, so that the result is analytic for low energy and
momenta. Also, $m_e$ provides the natural dimensions to
compensate for
those of the modified vacuum and, thus, sets the scale
for this effect to be physically observable.
All the above results are perturbative and should only be believed
when small. The problem is that they are extremely small since any
easily accessible values of $\vec B$, $G_N (\rho+p)$, $T$ or $a^{-1}$
are negligible when compared to $m_e$. In other words, electrons are
too heavy in this context.

A second relevant observation is that the modification of the
polarization--averaged speed of light is always proportional to the
numerical factor 11. This was already pointed out with respect to
Eqs. \erolf,
\escharnhorst\ in Ref. \barton
 , where the author noted that this
factor is common to all coherent light-by-light scattering
\berestetskii.
 Notice,
however, that the same factor appears in Eq. \edrummond\ which is
unrelated to light-by-light scattering, hinting at some deeper
interpretation of the above results.

It is, nevertheless, hardly arguable that the most enticing feature of
the above results is the sign of the modification of the speed of light:
some vacua lead to $v<c$, some to $v>c$. This last
possibility has motivated further study \barton \daniels\ben\
\bartonscharnhorst. The reader will find there a controversial discussion
of phase, group and causal signals and, in particular,
of the effects  of $v>c$ on causality. Let us just say that,
in accordance with ref. \drummond, we find no grounds for violation
of  microcausality.

The purpose of our work is to deepen  and complete our understanding of
the modifications of the speed of light. We do this by
proceeding from the
more physical to the more field-theoretical aspects of the problem.
We first study the already known results presented above. We will give a
unified and very physical interpretation of them which explains in an
extremely simple manner the sign of the variation of the speed of light.
We then concentrate on the thermal vacuum and consider the modification
of
the speed of (high and low energy) massless particles for different
theories up to two loops. This will require a precise definition of what
is meant by speed of massless particles which is presented in Sect. 3.
A comprehensive
 one-loop study shows that field theories fall into two
categories for the purpose of this study: those with an explicit mass,
$m$, and those with an intrinsic mass scale, $\Lambda$. This is done
in Sect. 4. In Sect. 5, the two-loop study for a theory of the first
category, QED, is performed. This reproduces Eq. \erolf\
 at low energies
and leads to a new result at high energies. The two-loop study for a
theory of the second category, massless $g \phi^4$, is performed in
Sect. 6. This is mainly of field theoretical interest as in the Standard
Model only the quarkless QCD sector has no explicit mass. Confinement
only allows massless particles beyond the deconfinement temperature
which does not correspond to the low temperature cold heat bath
situation we are considering here ($T\ll m$ or $T\ll \Lambda$).
Still, it is not excluded that these results will be of some relevance
to  photon propagation if almost massless  pseudogoldstone bosons,
like invisible axions, exist. In Sect. 7, a comparison of the thermal and
Casimir vacua is performed. This allows us, first, to better understand
the meaning of a modified vacuum and, second, to introduce a unified
formalism which explains, both mathematically and theoretically, the
different signs of the modifications of the speed of light.
We finally summarize our most relevant results.
\bigskip
\newsec{The unified formula}
\bigskip

We mentioned in the previous section that all the
 polarization--averaged
modifications of the speed of light are proportional to the numerical
factor 11 and  inversely proportional to $m_e^4$. This hints at
the possibility of writing them in a unified form, which thus digs out
the common phenomenon underlying the physics behind them. This is indeed
so, as we now show.

Consider first the magnetic field vacuum. From Eq. (1.1) one finds the
averaged speed of light
\eqn\one{
v = {1 \over 4} \int_0^\pi d \theta\ \sin \theta\ \left(v_\parallel
 (\theta) +
v_\perp
(\theta)\right) = 1 - {22 \over 135}\
\alpha^2\ {\vec B^2 \over m_e^4}\,,}
          which, recalling that the energy density for a magnetic field
is given by
\eqn\two{
\rho_B = {\vec B^2 \over 2} > 0\,,}
can be written as
\eqn\three{
v = 1 - {44 \over 135}\ \alpha^2\ {\rho_B \over m_e^4}\ .}

For the Friedmann-Robertson-Walker vacuum,
 we recall that  a radiation
dominated universe, as corresponds to $T\ll m_e$,is described by
 $p = {\rho \over 3}$.
Moreover, the gravitational energy density is negative
\eqn\four{
\rho_G = - \rho < 0\ ,}
so that Eq.
 (1.2) can be written as
\eqn\five{
v = 1 - {44 \over 135}\ \alpha\ (G_N m_e^2)\ {\rho_G \over m_e^4}\ .}

On its own, the cold
 temperature vacuum is characterized by its Planck energy
density
\eqn\six{
\rho_T = {\pi^2  \over 15} T^4
 > 0\ ,}
so that Eq. (1.3) reads
\eqn\seven{
v = 1 - {44 \over 135}\ \alpha^2\ {\rho_T \over m_e^4}\ .}

Finally, the averaged speed of light for the Casimir vacuum is, from
Eq. (1.4),
\eqn\eight{
v = {1 \over 2} \int_0^\pi d \theta\ \sin \theta\ v(\theta) = 1 + {11
\pi^2 \over 24300}\ \alpha^2\ \left({1\over a\ m_e}\right)^4.}
Recalling the energy density of the Casimir vacuum \birrel,
\eqn\nine{
\rho_a = - {\pi^2 \over 720 a^4} < 0\ ,}
the speed of light can be written as
\eqn\ten{
v = 1 - {44 \over 135}\ \alpha^2\ {\rho_a \over m_e^4}\,.}

Note that
Eqs. \three, \seven\ and \ten\
are identical, and \five\
 also coincides with them if
we substitute $G_N m_e^2$ by $\alpha$.
 We therefore find the following remarkable result:
  the low energy modification
of the speed of light is proportional to the ratio of the
energy density of the
modified vacuum (as compared to the usual one) over $m_e^4$,
 with a universal numerical
coefficient and the corresponding coupling constants. If the
modified vacuum
has a higher energy density, then
 $v < c$. If it is lower, then
  $v > c$. This is
what explains the sign of the change of the speed of light.

The general equation we have found
 for the speed of light in a modified vacuum
allows us to make further predictions. As an instance, low--energy
photons should travel in an external electric field, $\vec E$, with
a speed
\eqn\extelectric{
v = 1 - {22 \over 135}\
\alpha^2\ {\vec E^2 \over m_e^4}\ .}
Another remarkable consequence of the general equation we have found
is the fact that only for an asymptotically flat universe
would the speed of light remain $c$. Matter, radiation and
gravitational energy  add to zero, which is the relevant physical
quantity that affects the velocity of low--energy photons at
this order of perturbation theory.

Let us finish this section mentioning one surprising feature. Notice
that the numerical coefficient is the same for the gravitational vacuum
as for the others, in spite of its being unrelated to light-by-light
scattering. Furthermore this factor 11 also appears in the coefficient
of the Euler-Poincar\'e characteristic spin $1 \over 2$ contribution to
the gravitational trace anomaly \birrel\pascual.
 These connections will be
discussed elsewhere.

\bigskip
\newsec{Definitions and framework}
\bigskip

Let us first define what we mean by speed of light, or better, speed of
a massless particle. Consider a massless scalar
particle which moves in a
thermal vacuum. Its propagator is
given by
\eqn\frameone{
{Z^{-1} \over q_0^2 - \vec {q^2} - \Pi (q_0^2 - \vec {q^2}) - \Pi_T
(q_0^2, \vec {q^2})} \ \ \ ,\ \ \ \Pi\ (0) = 0\ ,}
where $Z$ is the field renormalization constant, $\Pi$ the
zero temperature self-energy and $\Pi_T$ its temperature correction. The
separation of $q_0^2$ and $\vec {q^2}$ in $\Pi_T$ reflects the
Lorentz-symmetry breaking induced by the heat bath.

We will define the speed of a massless particle for real, rather
than virtual, particles. This
is because for gauge theories definitions on the pole, even at finite
temperature, ensure gauge invariance \pisarski.
 The energy-momentum
dispersion relation which characterizes the propagation of a real
particle is obtained from imposing that the denominator of Eq. (3.1)
vanishes. This leads to the implicit solution $q_0 = E (q),\ q \equiv
|\vec q|$. At one--loop order one finds an explicit solution independent
of $\Pi$,
\eqn\frametwo{
E^2 (q) = q^2 + \Pi_T^{(1)} (q^2, q^2)\ .}
Note that the standard vacuum polarization, $\Pi$, does not break
Lorentz invariance and, thus, does not change the energy--momentum
dispersion relation.

In general, perturbation theory allows to obtain an explicit solution
\eqn\framethree{
E^2 (q) = q^2 + f (q, T, m)\ ,}
where $f (q, T, m)$ depends on $\Pi$ and $\Pi_T$. Its
dependence on momentum, temperature and virtual particle mass has been
specified. If the theory is massless, $m = 0$, an intrinsic mass scale,
 $\Lambda$, will appear beyond one loop through the running
of the coupling constant. Then, $f (q, T, m)$ should be understood as $f
(q, T, \Lambda)$. Certainly the dependencies on $m$ and $\Lambda$ are
different.

The speed of the massless particle is then given (as for photons
traveling
through a medium that changes the relation between $E$ and $p$)
by
\eqn\framefour{
v(q) \equiv {\partial E \over \partial q} = {q + {1 \over 2} f' (q, T,
m) \over \sqrt{q^2 + f (q, T, m)}}\ \ \ , \ \ \ f' (q, T, m) \equiv
{\partial \over \partial q} f (q, T, m)\ .}
Expansions of potential relevance are
\eqn\framefive{
v \simeq 1 - {1 \over 2 q^2} \left(f(q, T, m) - q f' (q, T,
m)\right)\qquad  , \qquad  q^2 \gg f,\  q \gg f'\ ,}
with $v \simeq 1$,
\eqn\framesix{
v \simeq {f' (q, T, m) \over 2 \sqrt{f (q, T, m)}}\hskip 0.5cm , \hskip
0.5cm  f' \gg q,\ f \gg q^2\ ,}
and
\eqn\frameseven{
v \simeq {q \over \sqrt{f(q, T, m)}} \hskip 0.5cm , \hskip 0.5cm f' (q,
T, m) \ll q,\ f \gg q^2 \ ,}
 with $v \to 0$, in spite of going, to one loop, as the inverse
of the square root of the coupling constant.

Three particular cases will be of interest. First, let  $f$ be
constant in $q$,
\eqn\frameeight{
f (q, T, m) \equiv \lambda^2 (T, m)\ .}
 The whole effect of the heat bath is, then,  to give a mass to
the particle, which however remains free, see Eq. \three.
 The speed of the particle is
modified as corresponds to acquiring a mass: from Eq.
(3.5) and Eq. (3.7), as Eq.
(3.6) never applies,
\eqn\framenine{
\eqalign{
v \simeq 1 - {\lambda^2 (T, m) \over 2 q^2} \hskip 0.5cm &, \hskip 0.5cm
q \gg \lambda \hskip 5cm {\rm (a)}\cr
v \simeq {q \over \lambda (T, m)} \hskip 1cm &, \hskip 0.5cm \lambda \gg
q  \hskip 5cm {\rm (b)}\cr
}}

Second, let us consider the case
when no mass is induced at low energies,
\eqn\frameten{
\lim_{q \to 0} f(q, T, m) = 0\ .}
 Then, $v(q)$ is a genuine modification of the speed of the
massless particle at low energies, since it  remains massless. The
dispersion relation is usually non-free. If, furthermore,
\eqn\frameeleven{
f (q, T, m) \sim 0 (q^2) \qquad  , \qquad  q \ll T}
(recall that always $T \ll m$), then Eq. (3.5) always holds in
perturbation theory and
\eqn\frametwelve{
v (q \ll T) \simeq 1 + {f'' (0, T, m) \over 4} \hskip 0.5cm , \hskip
0.5cm f'' \ll 1\ .}
This is the situation which corresponds to the results
reviewed in the introduction. In the particular case $f (q, T, m) \sim
q^2$ for all momenta the dispersion relation would be free but the speed
of the massless particle would have changed according to Eq.
(3.12)  for
all values of $q$, {\sl i.e.}
 it would be a different constant now. We have
not encountered this situation in any of the theories we have
investigated.

Third, let us consider the case
when no mass is induced at high energies,
\eqn\framethirteen{
\lim_{q \to \infty} f (q, T, m) = 0\ .}
 Then, at high energies Eq. (3.5) holds and represents a genuine
modification of the speed of the massless particle at high energies,
for it   remains massless. The dispersion relation is never free.

We have been implicitly assuming that $f (q, T, m)$ is real. This is,
however, usually not so \weldon, but imaginary contributions of the
self energy are not relevant to the issue of the speed of the massless
particle. Also, no zero momentum  ambiguities \gribosky \bedaque\
 appear in
what follows.

For massless Dirac fermions the propagator is given by
\eqn\framefourteen{
{Z^{-1} \over q_0 \gamma_0 - \vec q \cdot \vec \gamma - \Sigma (q_0^2 -
{q}^2) (q_0 \gamma_0 - \vec q \cdot \vec \gamma) - \Sigma_{s, T}
(q_0^2, {q}^2) q_0 \gamma_0 + \Sigma_{v, T} (q_0^2, {q}^2)
\vec q \cdot \vec \gamma}\ ,}
 with a notation similar to the used in Eq. (3.1). Notice that no
temperature contribution of other Dirac matrix structures have been
included, because they will not appear for the theories and perturbative
order considered in this work. The dispersion relation is given by
\eqn\framefifteen{
q_0^2 \left(1 - \Sigma (q_0^2 - {q}^2) - \Sigma_{s, T} (q_0^2,
{q}^2)\right)^2 = {q}^2 \left(1 - \Sigma (q_0^2 -
{q^2}) - \Sigma_{v, T} (q_0^2, {q}^2)\right)^2\ .}
 At one--loop order this leads to
\eqn\framesixteen{
E^2 (q) = q^2 \left(1 + 2\ \Sigma_{s, T}^{(1)} (q^2, q^2) - 2\
\Sigma_{v, T}^{(1)} (q^2, q^2)\right)}
 or, with the notation of Eq. (3.3), to
\eqn\frameseventeen{
f (q, T, m) = 2 q^2 \left(\Sigma_{v, T}^{(1)} (q^2, q^2) - \Sigma_{v,
T}^{(1)} (q^2, q^2)\right)\ .}

For massless gauge bosons the corresponding analysis is slightly more
complicated \tarrach. The result to one--loop order is
\eqn\frameeighteen{
E^2 (q) = q^2 - {1 \over 2} (\delta^{ij} - \hat q^i \hat q^j)\ \Pi_{ij,
T}^{(1)} (q^2, q^2)\ ,}
 where $\hat q^i$ is the spatial unit vector and
  $\Pi_{ij, T}$ are the spatial components of the
temperature contribution to the self--energy
tensor $\Pi_{\mu v}$, so that
\eqn\framenineteen{
f (q, T, m) = - {1 \over 2} (\delta^{ij} - \hat q^i \hat q^j)\ \Pi_{ij,
T}^{(1)} (q^2, q^2)\ .}

The one--loop expressions also hold at, say, two loops when there are no
one--loop
 temperature corrections. This will happen for our two--loop QED
study, because we will always neglect the thermal contribution due to
massive virtual particles which are always exponentially suppressed by
a huge
Boltzmann factor. This is because we consider, as said in the
introduction, only cold heat baths, $T \ll m$.

Summing up: our test particle is real and massless and we only
thermalize
massless virtual particles, so that if the theory has a massive particle
it only enters in the
 form of a virtual non--thermalized contribution.

\bigskip

\newsec{One--loop results }
\bigskip

Let us consider the propagation of  massless  particles in a cold
heat bath, using perturbation theory at one loop.
 We are here not
concerned with hot heat baths, which require a different formalism
\braaten, because our initial motivation was the smallness
of the effects reviewed in the introduction, and their extension to high
energies and massless theories, but without changing the conditions
defining the vacuum. We will use the real time formalism \niemi,
 which
at the one--loop level reduces to the original one--component formalism
\dolan. Thus, a massless scalar boson propagator reads
\eqn\loopone{D_T (k, k')=\delta^{(4)}(k-k') \left[{i \over k^2+i\eta} +
{2 \pi \over e^{|k_0| \over T} -1} \delta (k^2)\right],\quad
 k^2 =
k_0^2-\vec k^2}
while a gauge boson propagator in the Feynman gauge is given by
\eqn\looptwo{D_T^{\mu v} (k, k') = - g^{\mu v} D_T (k, k')}
For a massless Dirac fermion one has
\eqn\loopthree{S_T (k, k') = \delta ^{(4)} (k-k') \ksl  \left[{i \over
k^2 + i \eta} - {2 \pi \over e^{|k_0| \over T} +1} \delta (k^2)\right]}
with $\ksl  \equiv k_\mu \gamma^{\mu}$. The above expressions separate
the
zero temperature from the finite temperature contributions. A heat bath
has, of course, other effects on the propagation of a particle (e.g.
diffusion), but the ones which cannot be told from the virtual particle
effects, because they conserve energy and momentum, are included in the
modified vacuum and are given by Eq. (4.1,2,3).

  Consider first massless $(m_e = 0)$ QED. For the photon one finds from
Eq. \framenineteen\
 and \loopthree\ that $f(q, T)$ is constant, {\sl i.e.}
Eq. \frameeight\ applies and
\eqn\loopfour{\lambda^2_\gamma (T) = 2 {e^2 T^2 \over 12}\ ,}
where we have stressed that any of the two virtual electrons
thermalizes, and $e$ is the electron charge. For the massless electron,
from Eq. \frameseventeen, \looptwo\ and \loopthree, again $f(q, T)$
comes out constant
\eqn\loopfive{\lambda_e^2 (T) = {e^2T^2 \over 6} + {e^2 T^2 \over 12}\ ,}
where we have separated the thermalized photon from the thermalized
electron contributions.
  The constancy of $f(q,T)$ at one loop turns out to be a general
   one--loop
 result for theories without explicit masses. Indeed,
 in $SU(N_c)$ Yang-Mills theory without
 quarks one obtains (ghosts are thermalized as scalar bosons)
\eqn\loopsix{\lambda_g^2 (T) = N_c {g^2_s T^2 \over 3}\ ,}
where $Nc$ is the number of colors and $g_s$ the strong coupling
constant. Also for massless ${g \over 4!} \phi^4$, where the tadpole is
the only diagram which appears at one loop, the result is constant
\eqn\loopseven{\lambda_\phi^2 (T) = {g T^2 \over 24}\ .}

  Consider now the other type of theories we are interested in: those
with a virtual non-thermalized massive particle. At one--loop
order  there are no thermal contributions in QED, as the test particle,
being massless, has to be the photon and the virtual ones are all massive
electrons. This will be important when QED is studied at two loops in
the next section. The
Weinberg-Salam model has, however, diagrams which contribute,
specifically the one with neutrinos both as test and thermalized
particles and the $Z^0$ as the virtual massive particle. The result is
\eqn\loopeight{f (q, T, M_z) = {g_W^2 \over 2} \left[{T^2 \over 6} +
{M_z^2 T \over 4 \pi^2 q} \int_0^{\infty} d x {1 \over e^x + 1} \ln
\left|{M_z^2 - 4q Tx \over M_z^2 + 4q Tx}\right|\right]}
where $g_W
 \equiv {e \over 2 \sin \theta_W \cos \theta_W}$, $\theta_W$ is
Weinberg's angle and $M_z$ is the $Z^0$ mass. This satisfies Eq.
\frameten\
but not Eq. \framethirteen. Thus, the neutrino remains massless
at low energies
and its low energy  velocity
 is modified according to Eq. \frametwelve,
\eqn\loopnine{v (q T \ll M_z^2) = 1 - g_W^2 {7 \pi^2 \over 45} \left({T
\over M_z}\right)^4 + {\cal O} \left({q^2 T^6 \over M_z^8}\right)\ ,}
which is always a small correction for cold heat baths, while at high
energies it goes like
\eqn\loopten{v (q T \gg M_z^2) = 1- g_W^2 {1 \over 24} \left({T \over
q}\right)^2 + {\cal O} \left({M_z^4 \over q^4}\right)\ ,\quad
 T \leq
q\ ,}
where we have specified the condition for perturbation theory to hold.
Notice that Eq. \loopten\ is like Eq. (3.9a) so that the
neutrino has
indeed acquired a mass at high energies. This can also be seen from the
fact that Eq. \loopten\ allows the limit $M_z \to 0$ to be taken,
without
being modified. But, as analyzed previously in this section, for $M_z =
0$ the whole temperature effect should give the neutrino a mass. Indeed,
putting $M_z = 0$ in Eq. \loopeight\ gives (recall that still only
neutrinos are thermalized)
\eqn\loopeleven{\lambda_\nu^2 (T) = g_W^2 {T^2 \over 12}\ ,}
so that Eq. \loopten\ can be written as
\eqn\looptwelve{v (q T \gg M_z^2) = 1 - {\lambda_\nu^2
 (T) \over 2 q^2}\ \
,\quad \lambda_\nu \ll q\ ,}
exactly as Eq. (3.9a). Thus, the neutrino does not
acquire a mass at low
energies, but it does so at high energies, and the mass it acquires is
precisely the one it would have gotten in the massless theory. In other
words
\eqn\loopthirteen{\lim_{q \to \infty} f(q, T, M_z) = f(q, T, 0) =
\lambda_\nu^2 (T)}
  The result Eq. \loopnine\
 (with $v < 1$ and its quartic $T$ dependence),
the result Eq.  \loopten\ with $v < 1$ and its quadratic $T$
 dependence)
and its interpretation Eq. \looptwelve\ as a mass effect are of general
validity at one loop. One can easily check them in other examples,
 {\sl e.g.} in an effective
theory of photons and neutral massive scalars or pseudoscalars.

  At one loop only two scenarios exist. In a massless theory the test
particle acquires a mass and nothing else. Its speed is modified
accordingly. In a massive theory the test particle remains massless at
low energies and its speed is slowed down as $\left({T \over
m}\right)^4$ at low energies. This can be interpreted as due to a change
in the vacuum energy density as done in section 2. At high energies,
where the mass of the theory should be irrelevant, it acquires a mass;
precisely the one it would have acquired in the massless theory. Its
speed is slowed down accordingly.

  From the results
   presented in this section, only the one corresponding to
the Weinberg-Salam theory can be straight-forwardly used for
phenomenological estimations. The corrections to the speed
of neutrinos
 are even smaller than the
ones of photons reviewed in the introduction.
\bigskip
\newsec{Two--loop QED}
\bigskip

We now turn to the study of the modification of the speed of
photons in QED at two loops. The calculation has already been
performed in the literature for low--energy photons \tarrach\scharnhorst.
 In such a case, the result can be obtained from the Euler--Heisenberg
energy density, which encodes the fermion box diagram for constant
external fields. At high energies, there is no possibility to
circumvent the use of the full machinery of quantum field theory
at finite temperature. Therefore, we proceed now to perform that
first principles computation.

  For QED, because in the limit in which only massless particles
thermalize $\Pi_{ij,T}^{(1)} = 0$, one can use at two loops all the
one--loop formulae and techniques. Thus from Eq. \frameeighteen\ and
Eq. \looptwo\ one obtains \tarrach
\eqn\qedone{f(q,T,m_e) = -i e^4\pi
 \int {d^4 k \over (2 \pi)^4}\ {1
\over e^{|k_0| \over T} - 1} \delta (k^2) A(q,k,m_e)}
where the integrand, which corresponds to the electron box loop,
is given by
\eqn\qedtwo{\eqalign{
 &A(q,k,m_e) = \int{d^n p \over (2 \pi)^n} \bigg[{4(n-2)^2 \over
D(0)D(0)}-{16(n-2) \over D(0)D(k)}-{2(n-2)(n+4) \over D(0)D(k+q)}\cr
+&{2(n-2)(n+8) \over D(0)D(q-k)}-{64\ q \cdot k \over D(0)D(q)D(k)}
-{32\ q
\cdot k \over D(0)D(k)D(q+k)}-{64\ m_e^2 \over D(0)D(0)D(k)}\cr
+&{64\ m_e^4
 \over D(k)D(q)D(k+q)}+{64\ m_e^4 \over D(0)D(0)D(k)D(q)}+{16  \
m_e^2 (5\ q \cdot k+2\ m_e^2) \over D(0)D(k)D(q)D(k+q)}\bigg]}}
where $n$ is the dimension of space-time and
\eqn\qedthree{D(k) \equiv (p-k)^2-m_e^2+i \eta}
The integral in Eq. \qedtwo\ is ultraviolet (UV) convergent. Using
Feynman parametrization, it is easy to carry out the fermionic momentum
integral as well as all the parametric ones but the last. This last
parametric integral would lead to complicate dilogarithmic expressions
and,
therefore, we prefer to give its result in terms of series,
\eqn\qedseries{\eqalign{
A(q,k,m_e)= {4 i \over  \pi^2}
 \sum_{l=1}^\infty
 {\Gamma^2(l)\over \Gamma(2 l+3)}&
\left[
 \left( 12 l^2+43 l^3+50 l^2+ 26 l-12\right)\right.\cr &\left.
-\left( (-1)^l+1\right)
\left( 16 l^4+73 l^3+106 l^2+64 l+24\right)\right]
  \left( {2 q k\over m_e^2}
\right)^l \ ,\cr}}
where $\Gamma$ is the gamma function.
It is clear that the series readily produces, after  integration
upon the thermalized momenta, the low--energy expansion
\eqn\qedfour{\eqalign{
f(q T\ll m_e^2) &= - {e^4 T^2 \over  \pi^4}
 \sum_{n=1}^\infty
{\Gamma^2(2n) \Gamma(2 n+1)
\zeta(2 n+2)\over \Gamma(4 n+1)}
{4 n^3+12 n^2+5 n+1\over(n+1)}
\left({4 q T\over m_e^2}\right)^{2 n}\cr
 &= -e^4 \left(
 {11 \over 4050}  {q^2 T^4 \over
m_e^4}+ {208 \pi^2\over 212625}
  {q^4T^6 \over m_e^8}+\dots\right) ,}}
where $\zeta$ stands for the Riemann zeta--function.
This gives the expression Eq. \erolf\ for the leading modification
of the speed of low--energy photons.
 Further higher order
terms can be obtained without trouble. For arbitrary
$q$ the computation is much more involved since we need to sum
the above series \bartonscharnhorst.
 This is sketched in the Appendix, the basic
idea being to go back to \qedseries\ and
subdivide it into simpler series which
can be  summed using logs, dilogs and trilogs.
 The result leads to a high--energy expansion whose first term is
\eqn\qedfive{f(qT\gg m_e^2)={1
 \over 48 \pi^2} e^4 T^2 \ln^2 {qT \over
m_e^2}+{\cal O} \left(T^2 \ln {qT \over m_e^2}\right)\ ,}
leading  to
\eqn\qedsix{v(qT\gg m_e^2)=
1-{\alpha^2 \over 6} \left({T \over q}\right)^2
\ln^2 {qT \over m_e^2}\ \ ,\ \ T \leq q\ .}
In the process of obtaining this limit, a non--trivial cancellation of
$\ln ^3$  takes place.

   The summary of the above computation is that
low--energy corrections are  of the same type as already obtained
at one loop order (cf. Eq. \loopnine), but this is not so at high
energies. Indeed Eq. \qedsix\ has an IR-sensitive $\ln^2$ factor
which is
not present in Eq. \loopten. The electron mass plays the role of an IR
regulator and the limit $m_e \to 0$ cannot be taken. This is consistent
with the fact that the $m_e \to 0$ limit is incompatible with the
approximations used (which led to $\Pi_{ij, T}^{(1)} = 0$). The massless
theory cannot be obtained from our results and has to be worked out
separately. We will do so in the next section for a simpler theory.
  Notice that Eq. \qedfive\ does not satisfy Eq. \framethirteen\ so that
one
cannot interpret it as a genuine modification of the speed of light but
rather to one corresponding to an induced high--energy mass which runs
according to
\eqn\qedseven{\lambda^2 (qT\gg m_e^2)
 = {\alpha^2 \over 3} T^2 \ln^2 {qT
\over m_e^2}\ .}
  The high energy correction Eq. \qedsix\ is never
  quantitatively more
relevant than the low--energy correction, Eq. \erolf, as
\eqn\qedeight{\left({T \over q}\right)^2 \ll
 \left({T \over m_e}\right)^4\
\ ,\ \ qT \gg m_e^2\ .}
  Massive theories do not offer better perspectives for obtaining
relevant modifications of the speed of light at high energies than at
low energies. Furthermore only at low
energies is the modification  genuine,
{\sl i.e.} unrelated to an induced mass.
\bigskip
\newsec{Two--loop massless $g \phi^4$}
\bigskip

The results we have obtained so far stress
 the relevant role played
by masses in thermal corrections. For QED, $m_e$ has either
compensated for the dimensions carried by the temperature or played
the role of an IR cutoff.
 It is,
therefore, natural
 to ask the question of what would happen in a theory
without any extrinsic
 mass. This turns out to be a subtle problem at two
loops, that
we will study in the  laboratory of
   massless
   ${g\over 4\!} \phi^4$,
which is substantially simpler than massless QED.

Let us first recall the one--loop
 results, coming from  the tadpole diagram
\eqn\gphione{\eqalign{
&\Pi^{(1)} = 0\ ,\cr
&\Pi^{(1)}_T = {g T^2 \over 24}\ ,\cr}}
where we have used dimensional regularization, which preserves at zero
temperature the masslessness of the theory. At two loops, Umezawa's two
component formalism has to be used \niemi.
 This means that Eq. \loopone\ is generalized to
\eqn\gphitwo{D_T(k,k')=\delta^{(4)} (k-k') \left[{i \over k^2+i \eta\
(^1\ _{-1})} \bigg({1 \atop 0   }{0    \atop
 -1}\bigg) + {2 \pi \delta(k^2) \over
e^{{|k_0| \over T}} -1} \left({1
\atop e^{{|k_0| \over 2T}}} {e^{{|k_0| \over 2T}} \atop 1}\right)
\right]\ ,}
Fields are accordingly doubled so that
 the upper component is the physical field and the lower, the
auxiliary (and always virtual) field which only couples to itself with a
quartic coupling of opposite sign. Therefore,
 physical fields only connect to
auxiliary ones through the off--diagonal thermal propagator.
  The zero temperature two--loop contributions are
\eqn\gphithree{\Pi^{(2, t)} = 0\ ,}
for the double tadpole, and $(n=4+2 \epsilon)$
\eqn\gphifour{\Pi^{(2, s)} (q_0^2-q^2)={g^2 \over 24(16 \pi^2)^2}
\left({1 \over \epsilon} - \ell n {q_0^2-q^2 \over
\nu^2}+c_1\right)(q_0^2-q^2)\ ,}
for the setting sun diagram. The constant $c_1$ depends on the precise
definition of the regularization scale $\nu$. Eq. \gphifour\ of course
vanishes on the mass shell. Since Eq. \gphione\ is constant,
the two--loop generalization of Eq. \frametwo\ is
\eqn\gphifive{E^2=q^2+{g_0 T^2 \over
24}+\Pi_T^{(2,t)}+\Pi_T^{(2,s)}(q^2,q^2)\ ,}
where we have made explicit the bare character of the coupling constant
for the one--loop contribution.

  Consider now the different thermal contributions to the two--loop
tadpole diagram. With the help of Eq. \gphitwo, the two--point
 loop gives
a contribution proportional to
\eqn\gphisix{\eqalign{
 &\int{d^n k \over (2\pi)^n} \left[\left({i \over k^2+i\eta}+{2\pi
\delta(k^2) \over e^{{|k_0| \over T}}-1}\right)^2 - \left({2\pi
\delta(k^2) \over e^{{|k_0| \over T}}-1}\right)^2 e^{{|k_0| \over
T}}\right]\cr
=&\int{d^n k \over (2\pi)^n} \left[-\left({1 \over
k^2+i\eta}\right)^2 + {4 \pi i \delta(k^2) \over e^{{|k_0| \over T}} -1}
\left({1 \over k^2+i\eta}+ i \pi \delta (k^2)\right)\right]\ ,}}
so that, recalling
\eqn\gphiseven{\delta(k^2) P{1 \over k^2} = 0\ ,}
only the zero temperature contribution is left. Thus the whole thermal
contribution comes from the other loop. Note also that
 the Feynman integral
\eqn\gphieight{\int{d^n k \over (2\pi)^n}\ {1 \over (k^2+i\eta)^2}}
vanishes in dimensional regularization due to a cancellation of UV and
IR divergences. As temperature does not affect renormalizability we know
that Eq. \gphifive\ is UV-finite. It might, however, be IR-singular on
the grounds of being a massless theory which is, furthermore, considered
on the mass shell. It is, therefore, important to tell IR from UV
divergences. This we do by introducing where needed the one--loop
 thermal
mass as an IR regulator in the two--loop
 contributions. In particular, Eq.
\gphieight\ becomes
\eqn\gphinine{\int {d^n k \over (2\pi)^n}\ {1 \over \left(k^2-{g T^2
\over 24} + i\eta\right)^2} = {i \over (4\pi)^2} \left(- {1 \over
 \epsilon} -
\ln {g T^2 \over \nu^2} + c_2\right)\ .}
Here $c_2$ is a constant which depends on the precise value of the IR
regulator, which is actually arbitrary. Putting everything together, one
finds
\eqn\gphiten{\Pi_T^{(2,t)} = {g^2T^2 \over 48}\ {1 \over 16 \pi^2}
\left({1 \over \epsilon}
 + \ln {g T^2 \over \nu^2} - c_2\right)\ .}
It is worth mentioning here that both in Eq. \gphione\ and Eq.
\gphiten\
the thermal loop integral has been performed in $n=4$ dimensions. It
will immediately become clear why there is no need to dimensionally
regularize it.

  The thermalized setting sun diagram has two real contributions. One
corresponding to one thermalized particle. It is UV-divergent and
IR-finite:
\eqn\gphieleven{\Pi_T^{(2,s,1)} (q^2,q^2) = {g^2T^2 \over 24}\ {1 \over
16 \pi^2} \left({1 \over \epsilon}
 + \ln {qT \over \nu^2} + c_3\right)\ .}
Again, the thermal loop integral has been performed in $n=4$ dimensions.
  The other contribution, corresponding to two thermalized particles, is
both UV and IR-finite. It is proportional to the integral
\eqn\gphitwelve{\eqalign{
I(q,T) \equiv\ &g^2T^2 \int_0^\infty d x {x \over e^x -1}\ \int_0^\infty
d y {y \over e^y -1}\ \int_{-1}^1 d\mu\ \int_{-1}^1 d\rho
\sum_{\eta=\pm 1} \sum_{\eta'=\pm 1}\cr
&{1 \over {q \over T} (\eta x + \eta'x + \mu x) + \eta
\eta' xy + y \rho\ \sqrt{({q \over T})^2 + x^2 - 2 {q \over
T} x \mu} \ ,}\cr}}
which reduces to
\eqn\gphithirteen{\eqalign{
I(q, T) =& {4g^2T^3 \over q} \int_0^\infty d x\
\bigg[\bigg({q \over T} + x\bigg) \ln \bigg|{q \over T} +
x\bigg|
+ \bigg({q \over T} - x\bigg) \ln \bigg|{q \over T} - x\bigg|\bigg]
\cr  &\bigg(\ln (1-e^{-x})
- {x \over 2} {1 \over e^x-1}\bigg)\ .\cr}}
The high--energy and low--energy limits of $I(q, T)$ are
\eqn\gphifourteen{I(q\gg T)
 \simeq - {2 \pi^4 \over 9} g^2 {T^4 \over q^2}
+ {\cal O}\bigg({T^6 \over q^4}\bigg)}
and
\eqn\gphififteen{I(q\ll T) \simeq 4g^2T^2 \left[3 \sum_{n=1}^\infty {\ln
 n \over n^2} - \bigg(2 - {3 \gamma \over 2}\bigg) {\pi^2 \over
3}\right]\ ,}
where $\gamma$ is Euler's constant. Putting together Eq. \gphiten,
\gphieleven\ and
\eqn\gphisixteen{\Pi_T^{(2,s,2)} (q^2, q^2) = {1 \over 16 (2 \pi)^4}\
I(q, T)\ ,}
we get, from Eq. \gphifive,
\eqn\gphiseventeen{E^2 = q^2 + {g_0 T^2 \over 24} + {g^2 T^2 \over 48}\
{1 \over 16 \pi^2} \bigg({3 \over \epsilon}
 + \ln {q T^2 \over \nu^2} + 2
\ln {q T \over \nu^2} + c \bigg) + {1 \over 16(2\pi)^4}\ I(q, T)\ .}
Recalling the coupling constant renormalization
\eqn\gphieighteen{g_0 = g(\mu^2) \left(1 - {3g \over 32 \pi^2} \bigg({1
\over \epsilon}+
 \ln {\mu^2 \over \nu^2} + {c \over 3}\bigg)\right)\ ,}
where $\mu$ is the renormalization point, we find
\eqn\gphinineteen{E^2 = q^2 + {g(\mu^2)T^2 \over 24} + {g^2T^2 \over 768
\pi^2} \bigg(\ln   {g T^2 \over \mu^2} + 2 \ln   {qT \over \mu^2}\bigg)
+ {1 \over 256 \pi^4} I(q, T)\ .}
Notice that,
 due to the high--energy and low--energy behavior of $I(q,T)$,
Eq. \gphifourteen\ and \gphififteen\ respectively, its contribution
to Eq. \gphinineteen\ can be neglected, as it is subdominant. It is now
also clear why in Eq. \gphifive, Eq. \gphiten\ and Eq. \gphieleven\
there was no need to dimensionally regularize the thermal Feynman
integral: as eventually all UV divergences cancel in Eq. \gphinineteen,
the $0(\epsilon)$ terms, which one can see that they factorize, do not
contribute.
  The $\mu$-dependence in Eq. \gphinineteen\ is of course fictitious, so
that we can immediately renormalization group improve it by substituting
$(g T^4 q^2)^{1/3}$ for $\mu^2$, so that finally (neglecting $I(q, T)$),
\eqn\gphitwenty{E^2 = q^2 + {g \left((g T^4 q^2)^{1/3}\right) T^2 \over
24}\ .}
Thus, the whole two--loop contribution is contained in the running of the
coupling constant which appears at one loop,
\eqn\gphitwentyone{g (\mu^2)
 = {g(\Lambda^2) \over 1 - {3g(\Lambda^2) \over
32 \pi^2} \ln  {\mu^2 \over \Lambda^2}}\ ,}
with $\Lambda$ the intrinsic mass scale of the theory. Eq. \gphitwenty\
with Eq. \gphitwentyone\ is the main result of this section. The whole
two--loop temperature effect is to make the one--loop
thermally--induced
mass run.
  For very high values of $q$,
\eqn\gphitwentytwo{(g\ T^4 q^2)^{1/3}\ {\leq}\ \Lambda^2 e^{{32
\pi^2 \over 3g (\Lambda^2)}}\ ,}
one starts being sensitive to the Landau pole and perturbation theory
breaks down. Below that,
\eqn\gphitwentythree{(g\ T^4 q^2)^{1/3} \simeq \Lambda^2\ ,}
the running is too little conspicuous to the noticed: the whole
temperature effect is a constant mass. At much lower values,
\eqn\gphitwentyfour{(g\ T^4 q^2)^{1/3} \ll \Lambda^2 e^{-{32 \pi^2 \over
3g(\Lambda^2)}}\ ,}
the running coupling vanishes logarithmically and
\eqn\gphitwentyfive{E(q) \simeq q - {2 \pi^2 T^2 \over 9 q\ \ln \ {(g
T^4 q^2)^{1/3} \over \Lambda^2}}\  ,\quad T \ll q\ ,}
where we have made explicit that still the conditions corresponding to
Eq. \framefive\ hold, i.e. the kinetic energy still dominates over the
mass. From Eq. \framefive, one finds
\eqn\gphitwentysix{v = 1 + {2 \pi^2T^2 \over 9 q^2 \ln
{(g T^4q^2)^{1/3} \over \Lambda^2}}\ \leq\ 1\  ,\quad  T\ll q\ ,}
which corresponds to the modification of the speed of a massless
particle which has acquired a logarithmically vanishing mass. As $q$
becomes even lower, $q\ll T$, an IR divergence appears in Eq.
\gphinineteen. It requires an IR regulator, which comes in both
recalling that propagators can be regulated with the one loop thermal
mass or recalling that due to the induced thermal mass the situation of
the pole is shifted. In any case, $q^2$ friezes out at $g T^2$ and we
find a constant low energy mass
\eqn\gphitwentyseven{\lambda^2 (q\ll T) = {4 \pi^2 T^2 \over 9 \ln
  {g^{2/3} T^2 \over \Lambda^2}}\ ,}
with the corresponding non-relativistic velocity, Eq. \frameseven.

  It is worth stressing again that Eqs. \gphinineteen\ or \gphitwenty\
are IR sensitive results, as also
happened, somewhat differently, for the two loop QED results.
  For $g \phi^4$ the two--loop temperature effects are basically absorbed
by the running of the one--loop thermal mass, which does so according to
the zero temperature one--loop running of the coupling constant. There
 is, thus,
 no genuine modification of the speed of the massless particle. We
surmise that this result is of some generality, at least for theories
like massless QED which have the same asymptotic behavior. They might
hold, mutatis mutandis, for asymptotically free theories like pure QCD,
although more serious IR problems should be expected for them.
\bigskip
\newsec{Casimir vs. temperature}
\bigskip
  We now turn to the study of the propagation of photons in a Casimir
vacuum, defined by the presence of two infinite, ideal conducting
plates. The modifications of observable quantities are ultimately
related to the careful implementation of boundary conditions. In a
way, the same is true for thermal effects. There, periodic boundary
conditions are imposed on imaginary time. We shall keep this
analogy present all along our presentation.

  Let us start by comparing
   Eq. \erolf\ ( propagation in a heat bath) with Eq.
\escharnhorst\ ( propagation perpendicular to the Casimir plates,
$\cos \theta = 1$). Clearly both expressions are identical with the
substitution $2T \to a^{-1}$, but for a sign.
This is not surprising as temperature corresponds in the imaginary time
formalism to periodic boundary conditions and Casimir plates to
 half--periodic
 ones in one space direction. As said above, one expects a correspondence
between the formalisms which makes this identity apparent.
 This was already
pointed out by Barton \barton. In the same spirit but within a different
formalism we want to obtain the analog to Eq. \looptwo\ for the Casimir
vacuum. This will also allow us to gain more understanding of what a
modified vacuum is.

  The photon propagator for the Casimir boundary conditions $E_1 = E_2
=B_3 = 0$ on $x_3 = \pm {a \over 2}$ has been given in ref.
\bordag. It has the form
\eqn\casimirone{\bar D_c^{\mu \nu} (k, k') = \delta^{(4)} (k-k') D^{\mu
\nu}(k) + \delta^{(3)} (\tilde k - \tilde {k'})D_a^{\mu \nu} (\tilde k;
k_3, k'_3)\ ,}
where $D^{\mu \nu}(k)$ is the standard free propagator and
 $D_a^{\mu
\nu} (\tilde k; k_3, k'_3)$ is the gauge--independent modification due to
the boundaries. The actual form for this second piece of the propagator
was gotten in ref. \bordag\ in position space
\eqn\casimirprop{\eqalign{
 D^{\mu\nu}_a(\tilde x-\tilde y&; x_3,y_3) =
 {1\over 4} \int
 {d^3 \tilde k\over (2 \pi)^3} e^{-i \tilde k\cdot (\tilde x
 -\tilde y)} \left(-{\tilde g}^{\mu\nu}+{\tilde k^\mu
 \tilde k^\nu \over \Gamma^2(\tilde k)}\right)
  {1\over  \Gamma(\tilde
k) \sin(a\Gamma(\tilde k))} \cr  &
\left(
 e^{-i \Gamma(\tilde k) |x_3+y_3|}-
 e^{i \Gamma(\tilde k) |a-x_3+y_3|}-
 e^{i \Gamma(\tilde k) |a+x_3-y_3|}+
 e^{i \Gamma(\tilde k) |x_3+y_3|}   \right)\ ,}}
where
$\Gamma(\tilde k)={\tilde k}^2+i \eta$. Since
 $\tilde k \equiv (k_0, k_1, k_2,\ 0)$, it is clear that
momentum is not conserved in the 3-direction, as corresponds to the
breaking of translational invariance in this direction. On the other
hand,
 no
energy conservation violation takes place
 at finite temperatures, because
the boundary conditions are imposed in imaginary, not real, time.

Note that
Eq. \casimirone\ includes all the effects due to reflection at the
boundaries. As mentioned for finite temperatures after Eq. \loopthree,
one includes in the modified vacuum only those effects on the
propagation which cannot meaningfully be
 separated from the standard vacuum
propagation, { \sl i.e.}
 those contributions which also conserve energy and
momentum. For a Casimir vacuum this means studying the propagation far
from the plates. This implies to consider a wave packet localized in a
region far from the boundaries so that the limit $ka \to \infty$ should
be taken. From the above expression for the propagator,
this limit can be taken, obtaining
\eqn\casimirtwo{D_a^{\mu \nu} (\tilde k, k_3, k'_3)\ \sim
\ \delta(k^2)
\left[\delta(k_3+k'_3)- e^{i \Gamma(\tilde k)a} \delta(k_3-k'_3)
\right],\quad k a\to \infty\ .}
 In the above limit only two directions are
coherently enhanced, one of them forward and which thus conserves
momentum. This is the contribution which one cannot separate from
$D^{\mu \nu}$ and is, then, included into the Casimir vacuum structure.
This leads to the operative propagator
\eqn\casimirthree{D_c^{\mu \nu} (k,k') = \delta^{(4)} (k-k') \bigg[-
{ig^{\mu
\nu} \over k^2 + i \epsilon}
 + 2 \pi \left({\tilde g^{\mu \nu}} - {\tilde
k^\mu \tilde k^\nu \over \Gamma^2(\tilde k)}\right) {\delta(k^2)
\over e^{-2i \Gamma(\tilde k)a} -1}\bigg]}
where $D^{\mu \nu}(k)$ has been written in the Feynman gauge and
${\tilde g^{\mu \nu}} = g^{\mu \nu}+\delta^{\mu 3} \delta^{\nu 3}$.

 The
similarity of Eq. \casimirthree\ with Eq. \loopone,  with the
substitution $2T \to a^{-1}$, catches the eye, except for the fact that
in Eq. \casimirthree\ the exponential has an imaginary exponent and that
the tensor in the $a$-dependent part of Eq. \casimirthree\ is a
projector over physical helicities only in the 3-direction. This
explains the $\cos \theta$ dependence of Eq. \escharnhorst, while the
imaginary exponent implies a Wick rotation when the loop integral is
performed in the Casimir case which did not happen for the temperature
vacuum, leading to an analytic continuation $q^2 \to -q^2$. Our
results Eq. \qedfour\ and Eq. \qedfive\ would, then, become
\eqn\casimirfour{f({q \over a}\ll m_e^2)
 = {11 \over 4050}\ e^4\ {q^2 \over
(a\ m_e)^4}}
and
\eqn\casimirfive{f({q \over a}\gg m_e^2)
 = {1 \over 48 \pi^2}\ e^4 T^2 \ln^2\ {q \over a\ m_e^2}\ ,}
leading to
\eqn\casimirsix{v({q \over a}\ll m_e^2) = 1+{11 \pi^2 \over 8100}\
\alpha^2\ \big({1 \over a\  m_e}\big)^4 \ > 1}
and
\eqn\casimirseven{v({q \over a}\gg
m_e^2) = 1- {\alpha^2 \over 6} \big({1
\over a\ q}\big)^2 \ln^2 {q \over a\  m_e^2}\  <1\ ,}
always
for propagation perpendicular to the plates.

 The low--energy
expression,
with $v>1$, has been interpreted in section 2 as due to a decrease in
the vacuum energy density. The high--energy expression, with $v<1$, has
been interpreted in section 5 as due to a high--energy induced mass
which, of course, always decreases the speed of light. Both results,
Eq. \casimirsix\ and Eq. \casimirseven\ required, as discussed before,
$a\ q \gg 1$ as well as $a\ m_e\gg 1$,
 the condition equivalent to
coldness  which, here, means that the electron Compton wave length
should be smaller than the distance between plates. These two
restrictions
not only do not contradict the inequalities that hold for Eq.
\casimirsix\ and Eq. \casimirseven\ but, rather, ensure that the
corrections are small so that perturbation theory holds.

  On the contrary, for a cold heat bath ($T \ll m_e$),
   the other condition,
$T\ll q$,
does not follow. This, again, shows that a boundary condition in
imaginary time does not impose kinematical constraints as does a
boundary condition in real space. In fact, $q\ll T$ was
 considered in Eq.
\gphitwentyone.

  Physically, $v>1$ in Eq. \casimirsix\ can be interpreted, in harmony
with the analysis of section 2, in terms of a decrease in the number of
modes, recalling that real boundary conditions suppress modes.
Instead, the presence of
temperature adds real particles which allow for coherent scattering
 and, thus, gives $v<1$. On the contrary,
the result $v<1$ in Eq. \casimirseven\ can be understood by recalling
that in the standard vacuum the contributions of the different modes
balance in such a way that photons remain massless but that now, by
suppressing some modes, this balance is broken and a high energy mass
appears. This certainly slows down the propagation of photons.

 It is worth mentioning that examples of repulsive Casimir
effects are known. However, a detail analysis \candelas\
show that the energy density remains lower and, consequently,
the speed of light is faster than $c$.

  For gravity, the result $v>1$ \drummond\ \daniels\ can  be
interpreted similarly as a decrease in the number of modes, and $v<1$ as
the appearance of a mass, following the same line of thought as done
above for the Casimir vacuum.

\bigskip
\newsec{Conclusions}
\bigskip

  We have studied, up to two loops, the modifications of the speed of
massless particles in a cold heat bath and in other modified vacua, for
massive and massless theories, for high and low energies. We have found
a unifying expression for the low energy change in the averaged speed of
a light valid for all vacua:
\eqn\conclone{v = 1-{44 \over 135}\ \alpha^2\ {\rho \over m_e^4}}
where $\rho$ is the energy density and where, if the vacuum is a
gravitational one, one $\alpha$ has to be substituted by $m_e^2 G_N$.
It follows automatically that
 if the vacuum has a lower energy density than the standard vacuum,
$\rho<0$ and $v>1$, and viceversa.

 An enigmatic consequence of assuming full generality of the equation
we have just discussed takes place in cosmology. There, open, closed
and critical universes are distinguished by their total energy content.
It follows from Eq. \conclone\ that the critical universe is
singularized as low--energy photons travel in it with speed $c$,
due to the cancellation of all vacuum energy contributions.

  At one loop the whole temperature effect in the case of a massless
theory is just to give a mass. For a massive theory it leads to a
genuine change in the speed of the particle at low energies, but
suppressed by $\big({T \over m}\big)^4$. It parallels the two--loop
results in Eq. \conclone. At high energies, it leads to a change in the
speed of the particle as caused by a thermal mass, which  only plays
a role at high energies. The effect is suppressed by $\big({T \over
q}\big)^2$.

  At two loops the whole temperature effect in the case of the massless
theory chosen, $g \phi^4$, is to let the thermally induced one--loop
 mass
run, through the running of the coupling constant. It is an
IR--sensitive
result which requires an IR cutoff, which we have taken to be the one
loop--thermal
 mass. All the changes in the speed of the massless particle
are  due to the now running thermal mass. For the massive theory
chosen, QED, the two--loop low--energy result reproduces the known
modifications, and is suppressed by the same factor as at one loop. The
high--energy
 behavior, however, differs from the one obtained at one loop
by a squared logarithmic correction which, furthermore, depends on the
mass of the theory in an IR sensitive way. Nevertheless,
it can be
interpreted, like at one loop, as due to a now running thermal mass.

  Finally, the modification of the speed of light due to Casimir plates
has been compared and related to the one due to temperature.
A  field
theoretical explanation in terms of modes suggests the
following  physical picture
 of why photons move faster between plates than
in a normal vacuum, in contrast to what happens in a heated vacuum.
Modifications  of the vacuum that populate it with real or virtual
particles
introduce coherent scattering which reduces the speed of massless
particles. The opposite situation, {\sl e.g.} Casimir vacuum,
corresponds to modifications
of the vacuum that eliminate virtual modes and, consequently,
 their would--be scattering. The speed of massless particles is,
then, increased.

The most concise summary of our results states that
  only low--energy massless particles corresponding to a massive theory
show genuine modifications of their speed while remaining massless. All
other modifications are mass-related, or running mass-related.

\bigskip\bigskip
\noindent{\bf Acknowledgements}
\bigskip
We thank P. E. Haagensen and D. Z. Freedman for his comments
on the manuscript.  Financial support to this work has been
granted by CICYT, AEN-93-0695. Both R. T. and J. I. L.
acknowledge hospitality at CERN.

\vfil\eject
\noindent {\bf Appendix}
\bigskip

In section 5, we needed to calculate
\eqn\appone{f(q,T,m_e)= -{e^4\over 8\pi^4}\int_0^\infty {\rm d}k\
{k\over e^{k\over T}-1} R\left({ k q\over m_e^2}\right)\ ,}
where
\eqn\apptwo{
R(y) \equiv    \sum_{l=2}^\infty
{\Gamma^2 (l ) \over \Gamma(2 l +2)}2^{2 l+1}
 y^l  \big((-1)^l  +1\big)
\left(1+{2l ^2 \over (l +1)^2 (l +2)}\right)\ .}
First, we define
\eqn\appthree{R(y) = S(y) + S(-y)\ ,}
with $S$ containing all the powers in the series
\eqn\appfour{
S(y)  \equiv  \sum_{l =2}^\infty
{\Gamma^2 (l ) \over \Gamma(2 l +2)} 2^{2 l+1}
 y^l  \big(1+{2l ^2 \over
(l +1)^2 (l +2)}\big)\ .}
This sum can be splitted into three pieces,
\eqn\appfive{
S(y)= {1\over y^2} \left( S_1(y)+ 5 S_2(y) + 2 S_3(y)\right).}
 The first two are easily
carried out in terms of logarithms. The third requires changes
of variables to get it ready for
 resumation in terms of trilogarithms. These partial
results  are
\eqn\appsix{\eqalign{
S_1(y) &\equiv \sqrt{\pi} \sum_{n=4}^\infty {\Gamma^2(n-1) \over
\Gamma(n-{1\over 2})}\ {y^n \over n!}\cr
&= -y -y^2 -{2y^3 \over 9} - 2i \sqrt{y(1-y)} \ln (i \sqrt{y} +
\sqrt{1-y}) + (1-2y) \ln^2 (i \sqrt{y} + \sqrt{1-y})\ ,\cr
}}
\eqn\appseven{\eqalign{
S_2(y) &\equiv \sqrt{\pi} \sum_{n=4}^\infty {\Gamma(n-1) \Gamma(n-2)
\over \Gamma(n-{1\over 2})}\ {y^n \over n!}\cr
&= {y \over 2} + {7 \over 2} y^2 - {2 \over 9} y^3 + (1+2y) i
\sqrt{y(1-y)} \ln (i \sqrt{y} + \sqrt{1-y}) + (-{1 \over 2}+2y)
\ln^2 (i \sqrt{y}\cr
&+ \sqrt{1-y})\ ,\cr }}
\eqn\appeight{
\eqalign {
S_3(y) &\equiv \sqrt{\pi} \sum_{n=4}^\infty {\Gamma^2 (n-2) \over
\Gamma(n-{1\over 2}) (n-1)} {y^n \over n!}\cr
&= {y \over 4} - {53 \over 4} y^2 - {y^3 \over 9} + {i \over 2}
\sqrt{y(1-y)} (1-18y) \ln (i \sqrt{y}+\sqrt{1-y})\cr
&+ \big(- {1 \over 4} - 2y - 2y^2) \ln^2 (i\sqrt{y} + \sqrt{i-y}\big)
+ {4 \over 3}y \ln^3 (i\sqrt{y}+\sqrt{1-y})\cr
&+ 4y \ln^2 (i \sqrt{y}+\sqrt{1-y}) \ln (-2i \sqrt{1-y} \sqrt{y} +
2y)\cr
&- 4y \ln (i\sqrt{y} + \sqrt{1-y}) L_2\big[(\sqrt{1-y} +i \sqrt{y})^2
\big]\cr
&+ 2y L_3 \big[(\sqrt{1-y} + i\sqrt{y}\big] -2y \zeta (3)\ ,\cr}}
where we have used the notation
\eqn\appnine{
  L_n(y) = \sum_{k=1}^\infty {y^k \over
k^n}\ .}
Collecting this partial results we obtain
\eqn\appten{
\eqalign{
S(y) &= {1 \over  y^2} \bigg\{2y - 10 y^2 - {14 \over 9} y^3\cr
&+ (4 - 8y) i \sqrt{y(1-y)} \ln (i\sqrt{y}+\sqrt{1-y})\cr
&+ {8 \over 3}y \ln^3 (i\sqrt{y} + \sqrt{1-y})\cr
&- 8y \ln^2 (i\sqrt{y} + \sqrt{1-y}) \ln (-2i \sqrt{y(1-y)}+2y)\cr
&- 8y \ln (i\sqrt{y} + \sqrt{1-y}) L_2
\big[(i\sqrt{y}+\sqrt{1-y})^2\big]\cr
&+ 4y L_3 \big[(i\sqrt{y}+\sqrt{1-y})^2\big] -4y \zeta (3)\bigg\}
\ .\cr
}}
It is then easy to compute the limits used in the text,
\eqn\appeleven{\eqalign{
R(y) &{\mathop\sim_{y\to\infty}}\  -2 \ln^2 y\cr
&{\mathop\sim_{y\to0}}\ \ {88 \over 135} y^2 + {416
\over 3375} y^4 + \dots\cr} }

\vfil\eject
\listrefs
\end